# The laminar-to-turbulent transition in viscous fluid flow

**A. Paglietti**
*Department of Mechanical, Chemical and Materials Engineering
University of Cagliari, 09123 Cagliari, Italy
E-mail: paglietti@unica.it*

**ABSTRACT.** The onset of turbulence in laminar flow of viscous fluids is shown to be a consequence of the limited capacity of the fluid to withstand shear stress. This fact is exploited to predict the flow velocity at which laminar flow becomes turbulent and to calculate, on a theoretical basis, the corresponding critical value of the Reynolds number. A constitutive property essential to the present analysis is the ultimate shear stress of the fluid. The paper shows how this stress can be determined experimentally from a test in plane Couette flow. For water at 20 °C, the value of the ultimate shear stress is calculated from the experiments reported in the literature. This value is then is employed to predict the Reynolds number corresponding to the onset of turbulence in Taylor-Couette flow and in pipe flow of circular cross section. The results are realistic and their significance is assessed critically. The procedure can be applied to predict the onset of turbulence in any non-turbulent flow, provided that the velocity field of the flow is known.

## 1. Introduction

Laminar flow of viscous fluids becomes turbulent as the flow velocity exceeds a certain limit. The determination of this limit is a central problem of fluid mechanics with important implications for energy dissipation and fluid mixing. However, despite one and a half century of intense research since the pioneering work of Reynolds [20], the physical explanation of the origin of the laminar-to-turbulent transition is still missing, and there is a widespread consensus among physicists that the phenomenon is not fully understood (see, e.g., [24], [25] and [7] to quote a few references). The present paper provides the missing piece. The onset of turbulence is shown to be a necessary consequence of the angular momentum balance law and the limited capacity of a material to deform elastically without breaking.

In laminar flow, viscosity generates friction forces between adjacent laminae of fluid in relative sliding motion. To counterbalance the angular momentum that these forces apply to the elements of fluid, other shearing forces must be acting on the surfaces normal to the laminae. These shearing forces are not friction forces, because there is no sliding motion of fluid normal to the laminae. Instead, they arise from the mechanical stress produced by elastic deformation in the fluid. There is a limit, however, to the maximum stress that any material can oppose to elastic deformation without breaking. If the viscous forces applied to the lamina surface exceed the force that a fluid element can maintain without breaking, then the angular momentum balance cannot be met. The unbalanced angular momentum generates rotational motion in the elements of fluid, thus initiating the transition to a turbulent regime.

Two key ingredients of the analysis that follows are (*i*) the *elastic limit in shear* of the fluid and (*ii*) the *maximum shearing rate* at the points of the fluid. The elastic limit in shear is a constitutive property. It can be determined experimentally by testing the fluid in simple shearing (Sect. 4). On the contrary, the maximum shearing rate is independent of the fluid, but it depends on the particular flow that is being considered. The value of the maximum shearing rate at a point of the flow is determined by the velocity field at that point (Sect. 5). It can be calculated at any point of any flow, laminar or not, provided that the velocity field is well-defined and sufficiently smooth at the point. (These conditions are not met by turbulent flows, due to the almost random variations in the fluid particles velocity in turbulent regime.)

The above two ingredients are instrumental in predicting the critical velocity and, hence, the critical Reynolds number at which laminar flow starts becoming turbulent. The paper shows how this can be done in two examples that refer to water at 20 °C. For this medium the elastic limit in shear is first calculated from the experimental data on plane Couette flow available in the literature (Sect.4). This limit is then used to predict the critical velocity and the critical Reynolds number in the case of Taylor-Couette flow (Sect. 7) and in the case of pipe flow (Sect. 8). A similar analysis can be pursued for any flow, once the velocity field of the fluid is known, say from integration of the motion equations. The calculation of the maximum shearing rate at the points of the fluid can then be used to determine where and when turbulence is bound to appear in a flow. This opens new venues for drag-efficient design.

The topic considered in this paper has also been approached from a general thermodynamic standpoint in a recent book by this author [17]. The purely mechanical approach presented here, however, avoids the burden of thermodynamic arguments, still leading to general results that are directly applicable within the framework of classical fluid mechanics.

**2. Conflict between viscous forces and laminar flow**

When two contiguous parts of a body slide relative to one another, friction forces are produced on the contact surface to oppose the relative motion. This phenomenon takes place in every material, whether solid or fluid (liquid or gaseous). In fluids, friction occurs at interface between any two adjacent parts of fluid in relative motion. Fluid friction is usually referred to as viscosity and the forces that it produces are called viscous. However, we shall refer to these forces as "friction forces" whenever we want to emphasize that they originate from relative sliding of adjacent portions of material. In a laminar flow the fluid particles proceed in layers or laminae that slide over each other at different velocities without mixing. This produces friction forces on the surfaces of each lamina. The following arguments show that fluid friction tends to destroy rather than preserve the laminar character of the flow. Thus, to make laminar flow possible, other forces must be at work in the fluid to neutralize the effect of fluid friction.

Consider a plane laminar flow of a viscous fluid (Fig. 1). The flow is supposed to be steady, so that the velocity of the fluid particles does not change in time. Body forces are assumed to vanish. In the reference axes of Fig. 1, the velocity field is given by $\mathbf{v} = [v_1(x_2), 0, 0]$. Thus, every lamina of fluid parallel to the $(x_1, x_3)$-plane slides with respect to the adjacent laminae as represented in the figure. This generates friction forces at the lamina interfaces. These forces are applied to both surfaces of each lamina and are directed as the relative velocity of the adjacent lamina with respect to the considered one. Let's take a cubic element of fluid with sides parallel to the coordinate axes. Square *ABCD* in Fig. 1 gives a plane representation of such an element. Friction forces are applied to the faces of the cube where relative sliding of fluid takes place. These are faces *AB* and *CD*, normal to the $x_2$-axis and thus belonging to the surface of a lamina. If $t_f$ denotes the friction force per unit area acting on these faces, we have that $t_f = t_f(x_2)$. No friction acts on the other faces of the cube, as there is no sliding of fluid along them. In particular, no friction force is applied to faces *AC* and *BD* because they are orthogonal to **v**.

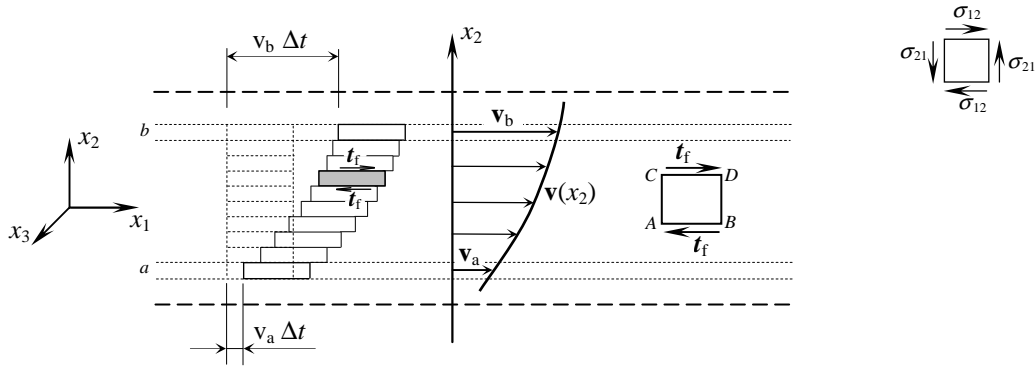

**Fig. 1.** Plane laminar velocity field $\mathbf{v} = \mathbf{v}(x_2)$ in a viscous fluid, showing the relative displacement between parallel laminae after time lapse $\Delta t$. Friction forces $t_f$ act per unit area on the surface of each lamina and, thus, on faces AB and CD of an infinitesimal cube parallel to the coordinate axes. *Inset*: Positive directions of stress components $\sigma_{21}$ and $\sigma_{12}$.

Pressure *p* (not shown in the figure) acts on all faces of the considered element. Thus the stress tensor at the element is given by

$$\boldsymbol{\sigma} = \begin{vmatrix} -p & t_f & 0 \\ 0 & -p & 0 \\ 0 & 0 & -p \end{vmatrix}, \qquad (2.1)$$

where $t_f$ is the component of $\mathbf{t}_f$ along the $x_1$-axis and the minus sign in front of the diagonal terms follows by taking *p* positive when representing compression. Result (2.1) can be proved easily by applying Cauchy's stress theorem,

$$\boldsymbol{t}^{(\mathbf{n})} = \boldsymbol{\sigma}\mathbf{n} \qquad or \qquad t_i^{(\mathbf{n})} = \sigma_{ij}\, n_j, \qquad (2.2)$$

to the case in which **n** coincides with the versors of the coordinate axes, namely $\boldsymbol{e}_1 = [1, 0, 0]$, $\boldsymbol{e}_2 = [0, 1, 0]$ and $\boldsymbol{e}_3 = [0, 0, 1]$, respectively. According to result (2.1) we have that

$$\sigma_{12} = t_f \qquad and \qquad \sigma_{21} = 0. \qquad (2.3)$$

That is, pressure and viscous forces do not suffice to meet the angular momentum balance at the considered fluid element, since that balance requires that the stress tensor should be symmetric. In other words, at each point of a fluid in laminar motion, fluid friction generates an unbalanced angular momentum, which, if not counteracted by other actions, would produce rotary motion in the fluid, thus making the considered laminar motion impossible. It must be concluded that other forces, in addition to pressure and friction, must be at work in order to make the stress tensor symmetric at every point of the fluid and thus the laminar flow possible.

**3. Elastic stress in a flowing fluid**

As is true of any real material, every fluid supports sound propagation. Therefore, to a greater or lesser extent all fluids must exhibit an elastic response to stress and strain. At a sufficiently small strain, the elastic stress-strain relation of an isotropic elastic material is given by the well-know relation of a linearly elastic isotropic medium, which can be expressed in the form [10]:

$$\boldsymbol{\sigma} = (K - \tfrac{2}{3}G)\operatorname{tr}(\boldsymbol{\varepsilon}^{\mathrm{e}})\boldsymbol{1} + 2G\,\boldsymbol{\varepsilon}^{\mathrm{e}} \qquad or \qquad \sigma_{ij} = (K - \tfrac{2}{3}G)\operatorname{tr}(\boldsymbol{\varepsilon}^{\mathrm{e}})\delta_{ij} + 2G\,\varepsilon_{ij}^{\mathrm{e}} \;. \tag{3.1}$$

In this relation, $\boldsymbol{\varepsilon}^{\mathrm{e}}$ denotes the elastic strain tensor (i.e., the elastic part of the total strain tensor), while $K$ and $G$ are the bulk modulus and the shear modulus of the material, respectively. These moduli are related to each other by the equation

$$G = \frac{3K(1-2\nu)}{2\,(1+\nu)} \tag{3.2}$$

where $\nu$ is the Poisson's ratio of the material. In a linearly elastic isotropic medium, the speed of sound or, more generally, the speed of pressure (or longitudinal) waves is given by

$$c_{\mathrm{L}} = \sqrt{\frac{1}{\rho}(K + \tfrac{3}{4}G)} = \sqrt{\frac{E(1-\nu)}{\rho(1+\nu)(1-2\nu)}} = \sqrt{\frac{2G(1-\nu)}{\rho(1-2\nu)}}\,, \tag{3.3}$$

where $\rho$ denotes mass density and $E$ is the modulus of elasticity of the material. In real materials, $c_{\mathrm{L}}$ is finite. This implies that $K$ should be finite and that $\nu < 0.5$, as evident from eqs. (3.3). In view of eq. (3.2), this also means that $G$ cannot vanish. It can be concluded that every fluid that supports pressure wave propagation at a finite speed must possess a non-vanishing elastic shear modulus.

Of course, this conclusion is quite another thing than stating that shear waves can propagate in fluids. Shear waves in fluids are not at issue here; although they can be detected in liquids [2], [3], [11], [12], [13], [15], and even in gases [6], and are predicted by the kinetic theory of gases [5], [14]. The difficulty with shear waves in fluids is that they can only propagate for such a short distance (often only a small fraction of a millimetre) that their occurrence is irrelevant for most practical purposes. However, the above conclusion concerning the value of $G$ is not related to shear wave propagation in fluids. That conclusion follows from the capacity of the fluid to support pressure wave propagation at a finite speed.

In order to better concentrate our attention on shear stress and shear strain, we shall henceforth assume that $p = 0$. This does not introduce any loss in generality, since pressure plays no role in the arguments that follow. Because $G$ is finite and different than zero, friction forces $t_f$ acting on faces $AB$ and $CD$ of the fluid element of Fig. 1 produce the elastic shearing strain given by:

$$\varepsilon_{12}^e = \frac{t_f}{2G} = \varepsilon_{21}^e. \tag{3.4}$$

As far as the present paper is concerned, the actual value of $G$ is immaterial, because the results of the analysis that follow do not depend on the value of $G$.

For the laminar motion considered in the previous section, it can be concluded from Eq. (3.4) that, in the considered reference system, the elastic strain tensor at each point of the fluid is given by:

$$\boldsymbol{\varepsilon}^e = \begin{vmatrix} 0 & \dfrac{t_f}{2G} & 0 \\ \dfrac{t_f}{2G} & 0 & 0 \\ 0 & 0 & 0 \end{vmatrix}. \tag{3.5}$$

Consequently, the corresponding stress tensor, as obtained from eq. (3.1), is:

$$\boldsymbol{\sigma} = \begin{vmatrix} 0 & \sigma_{12} & 0 \\ \sigma_{21} & 0 & 0 \\ 0 & 0 & 0 \end{vmatrix} = \begin{vmatrix} 0 & t_f & 0 \\ t_f & 0 & 0 \\ 0 & 0 & 0 \end{vmatrix}. \tag{3.6}$$

Therefore, the elastic deformation generated by the friction forces applied to faces *AB* and *CD* of the element produces symmetric stress components on faces *AC* and *BD* of the same element. These stress components enable the angular momentum balance to be met at the fluid elements. Thus the motion equations can be met and the considered laminar flow is possible.

**4. Elastic limit in fluids**

Real materials have a limited capacity to withstand stress without yielding or breaking. The stress limit beyond which a material ceases to respond elastically is a characteristic property of the material, whether solid or fluid, and is referred to as the material's elastic limit. Beyond the elastic limit, brittle materials break; other materials, mainly metals, deform permanently and may also harden before breaking. In general, the elastic limit is different for different states of stress (hydrostatic stress, uniaxial stress, pure shear stress, etc.). In this paper, we are exclusively concerned with the elastic limit in pure shear stress, or elastic limit in shear for short.

The elastic limit in shear is usually ignored in fluids because it is so small that it can be neglected in most of the applications. Fluids, moreover, have the remarkable capacity of repairing themselves immediately and seamlessly once broken apart (a common everyday experience, which repeats itself whenever any two parts of a fluid are separated from each other and then put back together again). This self-repairing capacity conceals the occurrence of rupture in a fluid, unless the rupture process does actually bring the broken pieces apart. For these reasons, the existence of an elastic limit is seldom if ever considered in fluids, in spite of the fact that all fluids, being capable of propagating sound waves, must possess elastic properties and hence also a limit to the elastic stress that they can exert.

Elastic limit to shear stress means, in particular, a limit to the maximum shear stress that the fluid can oppose to the applied friction forces. Since friction forces depend on the spatial derivatives of the velocity of the fluid particles, the elastic limit in shear entails a limit to the value of these derivatives and, thus, a limit to the velocity of any given laminar flow. This opens the way to relate the onset of turbulence to the value of the spatial derivatives of the velocity of the fluid particles. The details of the procedure are illustrated in Sect.6 with reference to linearly viscous fluids. Similar arguments can be pursued for any kind of fluid.

It seems reasonable to assume that in a fluid the stress at the elastic limit coincides with the ultimate stress that the fluid can oppose to elastic deformation. The experiments discussed below determine the ultimate shear stress of the fluid, which is, any way, the quantity that is needed in the analysis that follows. The question about the coincidence of elastic limit stress and ultimate stress is, therefore, of secondary importance.

The most direct way to determine the ultimate elastic shear stress of a fluid is to test it in homogeneous simple shearing flow *(plane Couette flow)*. An accurate instrument to do this is presented in [23]. In such experiment, the fluid is confined between two parallel plates that are kept in steady relative motion with respect to each other (Fig. 2). The lower plate is stationary, while the upper plate is driven at constant velocity $\bar{v}$ in the direction of the $x_1$-axis by a force $F$ acting in the same direction. Body forces are neglected because the distance $d$ between the plates is small. Pressure is constant throughout the fluid.

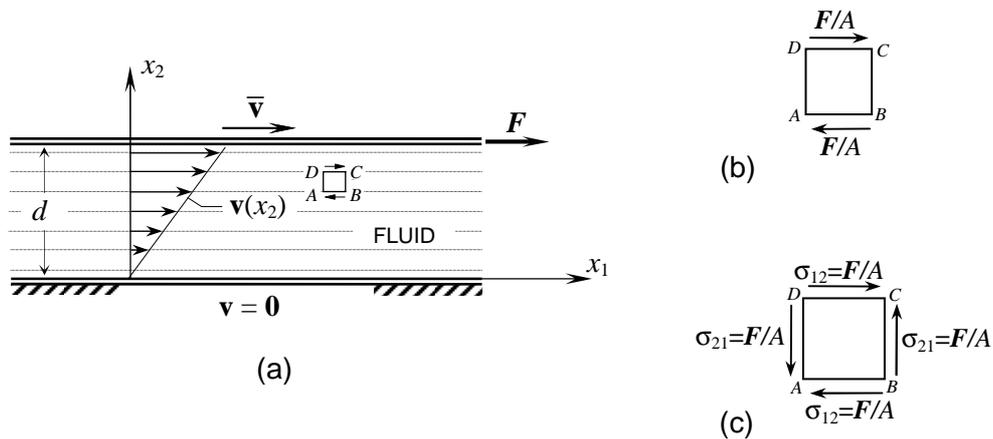

**Fig. 2.** (a) Homogeneous simple shearing flow between two parallel plates. (b) Viscous forces acting on fluid element *ABCD*. (c) Shearing stress component $\sigma_{21}$ needed to balance the element's angular momentum of the viscous forces.

In this flow, the fluid particles move along straight pathlines parallel to the $x_1$-axis. Their velocity **v** depends on the distance from the lower plate but is constant in each pathline. Adherence of the fluid to the plates is assumed. The motion equations require that, in steady state conditions, **v** should vary linearly from zero at the lower plate to $\bar{v}$ at the upper plate (Fig. 2). The velocity field of the fluid is, therefore,

$$v_1 = \frac{\bar{v}}{d} x_2, \quad v_2 = v_3 = 0, \tag{4.1}$$

$\bar{v}$ being the magnitude of $\bar{\mathbf{v}}$. As apparent from eq. (4.1), the single scalar $\bar{v}$ suffices to determine the velocity field at every point of the fluid. A major task of this experiment is to determine the tangential force $\boldsymbol{F}$ that is needed to keep the velocity of the upper plate constant. In linearly viscous fluids, $\boldsymbol{F}$ is found to be proportional to ratio $\bar{v}/d$ and, of course, to the area $A$ of the sliding plate. Therefore, on plane $x_2 = d$ the force that per unit area acting upon the fluid can be expressed as:

$$\frac{F}{A} = \eta \frac{\bar{v}}{d}, \tag{4.2}$$

where the proportionality factor $\eta$ is the viscosity coefficient of the material.

Because pressure is constant and body forces vanish, every plane normal to the $x_2$-axis exerts on the underlying fluid the same viscous force (4.2). The underlying fluid reacts by applying an equal and opposite force to the fluid above, as follows from Newton's third law. Accordingly, the viscous forces that are applied per unit area to the faces of a cubic element with sides parallel to the coordinate axes are the forces represented in Fig. 2b. As remarked in Sect. 3, to meet the angular momentum balance the element must deform elastically so as to develop stress component $\sigma_{21}=F/A$. This makes the stress tensor symmetric and, thus, the laminar flow (4.1) possible (Fig. 2c). From eq. (4.2) it follows, therefore, that at every point of the considered laminar flow the relation

$$\sigma_{21} = \sigma_{12} = \eta \frac{\bar{v}}{d} = \eta \frac{\partial v_1}{\partial x_2} \tag{4.3}$$

must apply. In the last of these equations we used the relation $\bar{v}/d = \partial v_1/\partial x_2$, which follows from eq. (4.1)$_1$ by taking the derivative with respect to $x_2$.

For the considered flow, the Reynolds number is usually expressed as:

$$R_e = \frac{\rho \bar{v} d}{\eta}. \tag{4.4}$$

Let $R_e^*$ be the value of $R_e$ at which the laminar to turbulent transition takes place. The corresponding velocity of the upper plate is denoted as $\bar{v}^*$. From eq. (4.4) we obtain:

$$\bar{v}^* = \frac{\eta R_e^*}{\rho d}. \tag{4.5}$$

This represents the maximum value of $\bar{v}$ that is compatible with the laminar flow.

Let $\tau_y$ denote the elastic limit in shear of the fluid. Because the laminar-to-turbulent transition occurs when the flow velocity exceeds the value at which $\sigma_{21}$ and $\sigma_{12}$ attain the elastic limit in shear, we conclude that $\sigma_{21} = \sigma_{12} = \tau_y$ for $\bar{v} = \bar{v}^*$. Therefore, from eq. (4.3) we infer that

$$\tau_y = \eta \frac{\bar{v}^*}{d}. \tag{4.6}$$

This formula can be used to determine the value of $\tau_y$ once $\bar{v}^*$ is obtained from the experiment.

Let's apply result (4.6) to determine the value of $\tau_y$ for water at 20 °C. According to the experiments of Tillmark and Alfredson [23], plane Couette flow of water at this temperature turns to turbulent at $R_e^* = 360$. The viscosity coefficient and the mass density of water at 20 °C are $\eta = 10^{-3}$ N sec/m$^2$ and $\rho = 10^3$ Kg/m$^3$, respectively. In the experimental apparatus adopted in [23] the depth of the flow was $d = 5$ mm. From this values and from eqs. (4.5) and (4.6) we therefore calculate that

$$\tau_y = 14.4 \; 10^{-3} \; [\text{Pa}]. \tag{4.7}$$

This is the ultimate elastic stress in shear of water at 20 °C. Being a property of the material, $\tau_y$ is independent of the particular experiment that is used for its determination.

For plane Couette flow of water at 20°C, the literature reports values of $R_e^*$ ranging from 280 to 750. The value of $\tau_y$ of water at that temperature may accordingly fall in the following range:

$$\tau_y = 11.0 \; 10^{-3} \div 30.0 \; 10^{-3} \; [\text{Pa}]. \tag{4.8}$$

In any case, such small values of $\tau_y$ explain why the elastic response of the fluid is ignored in fluid mechanics. Yet, $\tau_y$ dictates the transition from laminar to turbulent regime, because no laminar flow is possible as viscous friction exceed this limit. This fact opens the way to predicting the velocity at which a given laminar flow starts becoming turbulent and the corresponding value of $R_e$, once the value of $\tau_y$ of the fluid is known. The details of the procedure are illustrated in the following sections.

## 5. Maximum shearing rate at the points of a viscous flow

We know from classical fluid dynamics that the relative motion of fluid particles in the infinitesimal neighbourhood of a point is fully described by the rate-of-strain tensor *D* (cf., e.g., [4] and [1]). The components of this tensor in an orthogonal system of axes are determined by the velocity field of the fluid particles according to the relations:

$$D_{ij} = \frac{1}{2}\left(\frac{\partial v_i}{\partial x_j} + \frac{\partial v_j}{\partial x_i}\right). \tag{5.1}$$

The off-diagonal components of *D* represent rates of sharing on planes normal to the axes, while the diagonal components represent rates of stretching (or shrinking) along the axes. To be precise, let the components of *D* at a point *P* of the fluid be referred to an orthogonal system of axes (*x*, *y*, *z*). Let us then refer to component $D_{xy}$ of *D* and let us consider two infinitesimal material area elements about *P*, normal to the *y*-axis. The elements are assumed to be distant d*y* from each other. Component $D_{xy}$ can then be shown to equal the relative velocity in the *x*-direction of the two area elements, divided by d*y*. The same component is accordingly referred to as the rate of shearing at *P* in the x-direction. This interpretation extends immediately to the other off-diagonal components of *D*. A similar interpretation also applies to the diagonal components of *D* ([1], p. 212), but it will not be pursued any further, because it is not used in what follows.

As in any tensor, the components of *D* depend on the orientation of the reference axes. In view of the above interpretation of the off-diagonal components of *D*, it follows that the rate of shearing of parallel elements of area in the infinitesimal neighbourhood of a point is different for different orientations of the elements. Therefore, at every point of the fluid, there is an orientation of the elements, i.e., a direction of their normal, for which the rate of shearing is largest. The amplitude of the largest rate of shearing at the considered point will be denoted as $s_{\max}$ and it will be referred to as the *maximum shearing rate* at the point. Because *D* is a symmetric second order tensors, the maximum shearing rate at any point of the fluid can be determined by applying to *D* the same analysis that in the theory of elasticity is applied to stress tensor to determine the maximum shear stress component at a point of a body (cf., e.g., [22], [16] and [10]). Thus, if $D_1$, $D_2$ and $D_3$ denote the three principal values of *D*, assumed to be ordered in such a way that $D_1 \geq D_2 \geq D_3$, it is a straightforward matter to prove that

$$s_{\max} = \frac{D_1 - D_3}{2}. \tag{5.2}$$

Relation (5.2) enables us to calculate the maximum shearing rate at a point of the fluid once the flow velocity field, and hence tensor $\boldsymbol{D}$, is know at that point. In general, $s_{max} = s_{max}(x_1, x_2, x_3, t)$, because $\boldsymbol{D} = \boldsymbol{D}(x_1, x_2, x_3, t)$. Of course, in steady state conditions the dependence on time drops out from these functions.

As already observed, the relative sliding of adjacent portions of fluid generates friction forces on the sliding surface. The intensity of these forces depends on the relationship between friction and rate-of-strain. In linearly viscous fluids, friction equals the product of the relative shearing rate times the viscosity coefficient $\eta$ of the fluid. It follows that the amplitude $|t_f|_{max}$ of the largest friction force that per unit area acts at a point of a linearly viscous fluid can be obtained by multiplying by $\eta$ the maximum shearing rate at that point. That is:

$$|t_f|_{max} = \eta \, s_{max} . \tag{5.3}$$

In a laminar flow, the surfaces of maximum shearing rate coincide with the lamina surfaces. In plane laminar flows, moreover, the velocities of all the fluid particles are parallel to the plane of flow. In this case, the determination of the value of $|t_f|_{max}$ at a point $P$ of the fluid can be simplified by making reference to an orthogonal system of axes chosen as follows. Axis $x_1$ is directed as the fluid velocity at $P$; axis $x_2$ is normal to the lamina surface at $P$; while axis $x_3$ is, of necessity, normal to the plane of flow. In this system of axes, stress component $\sigma_{23}$ vanishes, so that the tangential component of stress vector on the lamina surface coincides with $\sigma_{21}$. By referring to that system of axes, we can therefore write:

$$|t_f|_{max} = |\sigma_{21}| . \tag{5.4}$$

To express this relation in terms of the derivatives of the velocity field at the considered point, we make recourse to the stress constitutive equation. For linear viscous fluids we have that:

$$\boldsymbol{\sigma} = -p\,\boldsymbol{1} + \lambda\,(\mathrm{tr}\,\boldsymbol{D})\,\boldsymbol{1} + 2\,\eta\,\boldsymbol{D} , \tag{5.5}$$

where $\lambda$ denotes the so called bulk viscosity coefficient. In view of eq. (5.1), equation (5.5) can be written as

$$\sigma_{ij} = -p\,\delta_{ij} + \lambda\,(\text{div }\mathbf{v})\,\delta_{ij} + \eta\left(\frac{\partial v_i}{\partial x_j} + \frac{\partial v_j}{\partial x_i}\right), \tag{5.6}$$

since $\text{tr }\mathbf{D} \equiv \text{div }\mathbf{v}$. The stress constitutive equations of incompressible linearly viscous fluids are recovered from the above equations by setting $\text{tr }\mathbf{D} = \text{div }\mathbf{v} = 0$. In view of eq. (5.6), relation (5.4) yields:

$$\left|t_f\right|_{\max} = \eta\left|\frac{\partial v_1}{\partial x_2} + \frac{\partial v_2}{\partial x_1}\right|. \tag{5.7}$$

On account of eq. (5.3), this result can also be written as:

$$s_{\max} = \left|\frac{\partial v_1}{\partial x_2} + \frac{\partial v_2}{\partial x_1}\right|. \tag{5.8}$$

Of course, eqs. (5.7) and (5.8) are only valid in the particular system of reference in which relation (5.4) holds true.

**6. Limit to the laminar flow**

The amplitude $|t_f|_{\max}$ of the largest friction force produced by a laminar flow at any given point of a linear viscous fluid can be calculated from eqs. (5.2) and (5.3) once the value of $\mathbf{D}$ at the point is known. As discussed in Sect. 4, the largest shear stress that a fluid can oppose to friction force cannot exceed the elastic limit $\tau_y$. If at a point of the fluid $|t_f|_{\max}$ exceeds $\tau_y$, the symmetry of the stress tensor breaks and, as a consequence, a local rotary motion is activated. This modifies the laminar velocity field of the original flow and produces its change to a turbulent regime. To be compatible with the motion equations and the elastic limit of the fluid, therefore, the laminar flow of a linear viscous fluid must meet the relation:

$$\left|t_f\right|_{\max} \leq \tau_y \tag{6.1}$$

at every point of the fluid. Accordingly, the relation:

$$\left|t_f\right|_{\max} = \tau_y \tag{6.2}$$

defines the limit of admissibility of the laminar flow.

By using eq. (5.3), condition (6.1) can be written as

$$s_{\max} \leq \frac{\tau_y}{\eta},\qquad(6.3)$$

which, in terms of the rate of strain tensor, becomes:

$$D_1 - D_3 \leq 2\frac{\tau_y}{\eta},\qquad(6.4)$$

as follows from eq. (5.2). In the particular cases in which eq. (5.8) applies, the above condition can be expressed in the form:

$$\left|\frac{\partial v_1}{\partial x_2} + \frac{\partial v_2}{\partial x_1}\right| \leq \frac{\tau_y}{\eta},\qquad(6.5)$$

which, when applicable, is often easier to use than eq. (6.4).

## 7. Example 1. Onset of turbulence in Taylor-Couette flow

Taylor-Couette flow refers to the steady-state flow of a viscous fluid confined between two coaxial cylinders that rotate at different angular velocities about their common axis. This axis is taken here as coinciding with the $x_3$-axis of the reference system. Some additional notation relevant to the present example is defined in Fig. 3. In laminar conditions, the fluid particles move in uniform circular motion in concentric circles on planes normal to the $x_3$-axis, centred on this axis. Accordingly, the flow is plane and the lamina surfaces are circular cylinders coaxial with $x_3$-axis. The angular velocity of the fluid particles is found to depend on the radius of their pathline according to the relation (cf., e.g., [4], [24] and [9]):

$$\omega = \omega(r) = -\frac{A}{2r^2} + B,\qquad(7.1)$$

where

$$A = \frac{2 r_1^2 r_2^2}{r_2^2 - r_1^2} (\omega_2 - \omega_1) \qquad \text{and} \qquad B = \frac{r_2^2 \omega_2 - r_1^2 \omega_1}{r_2^2 - r_1^2}. \qquad (7.2)$$

Uniform circular motion along the particle pathlines means that

$$\omega = \omega(r) = \frac{\mathrm{v}(r)}{r}, \qquad (7.3)$$

where $\mathrm{v} = \mathrm{v}(r)$ indicates the magnitude of the particle velocity. From eqs. (7.1) and (7.3), we obtain then

$$\mathrm{v} = \mathrm{v}(r) = -\frac{A}{2r} + B\, r. \qquad (7.4)$$

In the present case, the velocity vector **v** at any point $P$ of the fluid is normal to segment $OP$. Thus, on account of eq. (7.3), the components of **v** in the reference system of Fig. 3 are given by:

$$\mathrm{v}_1 = -\mathrm{v} \sin \omega t = -r \omega \sin \omega t = -\omega x_2, \qquad \mathrm{v}_2 = \mathrm{v} \cos \omega t = r \omega \cos \omega t = \omega x_1 \qquad \text{and} \qquad \mathrm{v}_3 = 0. \qquad (7.5)$$

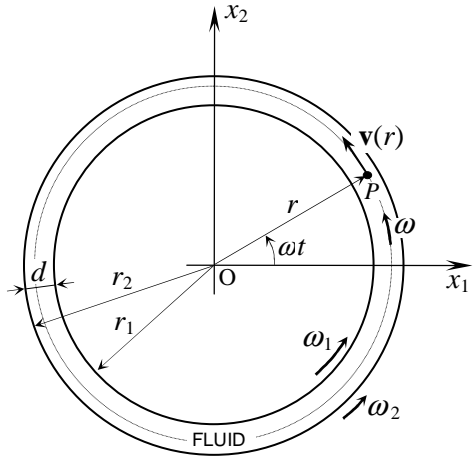

**Fig. 3**. Laminar flow of a fluid between two coaxial cylinders in relative rotation about their common axis (*Taylor-Couette flow*).

From the symmetry of the problem we can assume that the maximum shearing rate at a point of the flow –and thus the value of $|t_f|_{max}$ at that point– only depend on $r$. Therefore, in order to verify whether condition (6.1) is met at every point of the fluid, we can limit our considerations to the points on the $x_1$-axis. We can apply formula (6.5) to these points, since the $x_1$-axis is normal to the laminae, the $x_2$-axis is parallel to **v**, and the $x_3$-axis is normal to the plane of flow. On the $x_1$-axis we have that $x_1 \equiv r$ and $x_2 = 0$. Therefore, by taking the derivatives of eq. (7.3) along that axis we obtain $\partial \omega / \partial x_1 \equiv d\omega/dr$ and $\partial \omega / \partial x_2 = 0$. From eqs. (7.5) it can then been concluded that $\partial v_1/\partial x_2 = -\omega$ and that, moreover, $\partial v_2/\partial x_1 = (\partial \omega / \partial x_1 + \omega) \equiv (d\omega/dr + \omega)$. When applied to the present case, therefore, condition (6.5) becomes

$$\eta \, r \left| \frac{d\omega}{dr} \right| \leq \tau_y. \tag{7.6}$$

When the outer cylinder is fixed, we have that $\omega_2 = 0$. In this case, by means of eqs. (7.2)$_1$, (7.3)$_2$ and (7.4), we can write relation (7.6) as:

$$\omega_1 \leq \frac{r_2^2 - r_1^2}{2 \, r_1^2 \, r_2^2} \, \frac{\tau_y}{\eta} \, r^2. \tag{7.7}$$

The smallest value of the right hand side of this condition is attained for $r = r_1$, i.e., at the surface of the inner cylinder. Therefore, the largest value that $\omega_1$ can reach in laminar flow conditions is:

$$\omega_1^* = \frac{r_2^2 - r_1^2}{2 \, r_2^2} \, \frac{\tau_y}{\eta}. \tag{7.8}$$

For $\omega_1 > \omega_1^*$, turbulence starts to appear in the flow. The transition to turbulence is gradual, though, since the rate-of-strain is not uniform through the fluid. As $\omega_1$ exceeds $\omega_1^*$, turbulence is initially produced in a thin layer of fluid near the surface of the inner cylinder, since the right hand side of condition (7.7) is smallest there. Further increases in $\omega_1$ will extend the turbulent layer toward the outer cylinder until, eventually, it will involve the whole flow.

The Reynolds number associated with this flow is usually defined as

$$\mathrm{R_e} = \frac{\rho\,\omega_1\,r_1\,d}{\eta}. \tag{7.9}$$

Its critical value, $\mathrm{R_e^*}$, at the onset of turbulence is obtained from eq. (7.9) by substituting $\omega_1^*$ for $\omega_1$. By using eq. (7.8), we thus obtain:

$$\mathrm{R_e^*} = \frac{(r_2^2 - r_1^2)\,r_1\,d}{2\,r_2^2}\,\frac{\rho\,\tau_y}{\eta^2}. \tag{7.10}$$

Let us apply this result to predict the Reynolds number at the onset of turbulence for the Taylor-Couette flow considered in the experiments by Sinha *et al.* [21]. These experiments refer to water at room temperature, tested in an apparatus in which $r_1 = 3.81$ cm, $r_2 = 4.21$ cm, and $d = 0.4$ cm. In want of more precise data concerning the temperature of the experiments, we assume it to be 20 °C. Accordingly, we take $\rho = 10^3$ Kg/m$^3$, $\eta = 10^{-3}$ N sec/m$^2$ and, from our result (4.7), $\tau_y = 14.4\ 10^{-3}$ Pa. With these data, eq. (7.10) yields $\mathrm{R_e^*} = 197$. The experimental value determined in [21] is $\mathrm{R_e} = 239$ for $\omega_2 = 0$ and in the absence of axial flow. This value is some 20% larger than the value of $\mathrm{R_e^*}$ predicted by the present theory, although it is reasonably near to it in view of the uncertainties in the value of $\tau_y$. It should be observed, moreover, that the predicted value of $\mathrm{R_e^*}$ refers to the initiation of turbulence, whereas the experimental value quoted above refers to full-blown turbulence. Obviously, in a non-homogeneous flow, as the present one, full-blown turbulence is attained at a value of $\mathrm{R_e}$ somehow in excess of the threshold value $\mathrm{R_e^*}$ to which eq. (7.10) refers.

## 8. Example 2. Onset of turbulence in pipe flow

Steady laminar flow of incompressible linearly viscous fluids through rectilinear circular pipes is one of the best-known rigorous solutions of the motion equations of classical fluid mechanics. In the laminar regime at constant pressure gradient ($dp/dz = \Delta p/\Delta L = $ const), the fluid particles move at a constant speed along rectilinear paths, parallel to the pipe axis. The flow is also known as Poiseuille or Hagen-Poiseuille flow. In a cylindrical coordinate system ($r$, $\theta$, $z$) in which the inner wall of the pipe is $r = d/2$ and the $z$-axis coincides with the pipe axis, the velocity field of the flow is given by (cf., e.g., [3], [26]):

$$v_r = v_\theta = 0, \qquad v_z = v(r) = v_{max} - \frac{4 v_{max}}{d^2} r^2, \qquad (8.1)$$

where

$$v_{max} = \frac{1}{16 \eta} \frac{\Delta p}{\Delta L} d^2. \qquad (8.2)$$

The notation is that of Fig. 4.

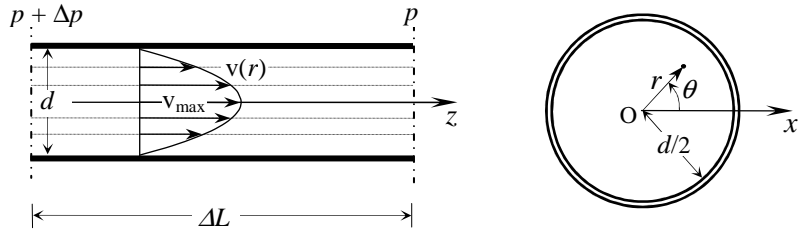

**Fig. 4.** Velocity profile of steady laminar flow of incompressible linearly viscous fluid in circular pipes (*Hagen-Poiseuille flow*).

The rate-of-deformation tensor, **D**, relevant to this flow is obtained by applying eq. (5.1) to the velocity field (8.1). In the considered coordinate system ($r$, $\theta$, $z$), the only non-vanishing components of **D** turn out to be $D_{rz} = D_{zr}$. That is:

$$\boldsymbol{D} = \begin{vmatrix} 0 & 0 & -8\dfrac{v_m}{d^2}r \\ 0 & 0 & 0 \\ -8\dfrac{v_m}{d^2}r & 0 & 0 \end{vmatrix}, \qquad (8.3)$$

where $v_m$ is the flow's mean velocity. In the present case:

$$v_m = \frac{1}{2} v_{max}, \tag{8.4}$$

as can be calculated from eq. (8.1). As apparent from eq. (8.3), the points of the fluid are subjected to a state of non-uniform shearing rate in simple shear. The shearing rate vanishes at the pipe axis ($r = 0$) and attains its largest value at the pipe wall ($r = d/2$).

In the present case, the principal values of $\boldsymbol{D}$ are:

$$D_1 = \frac{8\,v_m}{d^2} r, \qquad D_2 = 0, \qquad \text{and} \qquad D_3 = -\frac{8\,v_m}{d^2} r, \tag{8.5}$$

as can easily be determined either analytically or by applying Mohr's circle graphic representation [16]. The maximum shearing rate at the points of the fluid follows immediately from eq. (5.2):

$$s_{max} = \frac{8\,v_m}{d^2} r. \tag{8.6}$$

Therefore, from condition (6.3) or (6.4) we infer that the following inequality:

$$\frac{8\,v_m}{d^2} r \leq \frac{\tau_y}{\eta} \tag{8.7}$$

must be met at each point of the fluid for the considered laminar flow to be admissible. By taking the equality sign in this relation and by setting $r = d$, we calculate the maximum value of $v_m$ beyond which no full laminar flow is possible in the pipe:

$$v_m^* = \frac{d}{8} \frac{\tau_y}{\eta}. \tag{8.8}$$

The Reynolds number for fluid flow in circular pipes is usually defined as:

$$R_e = \frac{\rho\, v_m\, d}{\eta}. \tag{8.9}$$

By inserting $v_m^*$ for $v_m$ in this equation we obtain:

$$R_e^* = \frac{\rho v_m^* d}{\eta} = \frac{\rho \tau_y d^2}{8\eta^2}. \qquad (8.10)$$

This is the limit value of the Reynolds number above which full laminar flow through the pipe is not possible.

For example, for water at 20°C we have that $\rho = 10^3$ Kg/m$^3$, $\eta = 10^{-3}$ N sec/m$^2$ and $\tau_y = 14.4 \cdot 10^{-3}$ Pa, as recalled in the previous section. In this case, from eq. (8.10) we calculate

$$R_e^* = 180\, d^2 \qquad (d \text{ expressed in [cm]}). \qquad (8.11)$$

For $d = 3$ cm, this formula yields $R_e^* = 1620$, which is not an unreasonable value if compared with experiment.

No limit to laminar flow is predicted by classical fluid dynamics and Hagen-Poiseuille flow is known to be linearly stable for every value of $R_e$. Experiments show that the limit Reynolds number above which pipe flow is turbulent is usually between 2,000 and 4,000. There are experiments, however, of laminar flows in pipes at values of $R_e$ up to orders of magnitude greater than these (cf., e.g., [19], [18],[24]). Moreover, in pipes of very small hydraulic diameter, turbulence has been detected for values of $R_e$ as low as 200 - 400, cf. [8]. This large range of critical values of $R_e$ and the fact that, according to classical theory, pipe flow is unconditionally stable for every value of $R_e$ appear to indicate that the transition to turbulence in this flow is not fully controlled by the Reynolds number. As observed by Durst *et al*. [27]: "*No reason for this extended range is given in the literature. A closer look at existing data show, however, that there is a clear dependence of the critical Reynolds number on the employed pipe diameter.* […] *the existing data show an increase of the critical Reynolds number with increasing pipe diameter.*"

In the same paper [27] the dependence of the onset of turbulence on pipe diameter is ascribed to the effect of the shape of the nozzle at the pipe inlet. This is not inconsistent with the present findings. Different shapes of the inlet mean different velocity gradients and, thus, different values of shear stress at the pipe inlet. This makes the limit to laminar flow of the nozzle-pipe system depend on the nozzle shape. What all this means is that the transition to turbulence in pipe flow is not fully controlled by the Reynolds number.

In order to better understand why the critical value of the Reynolds number depends on $d$, it may help observe that it is true, of course, that for any given value of $R_e$ every pipe flow is governed by the same dimensionless equations, irrespective of the pipe diameter. However, the transition to turbulence is a different phenomenon. It depends on the elastic limit in shear of the fluid, which is a property that does not enter the motion equations. As a consequence, the onset of turbulence is not fully controlled by the Reynolds number. In pipe flow, the largest value of the maximum shearing rate is attained at the pipe wall ($r = d/2$) and it decreases as $d$ is increased, as evident by setting $r = d/2$ in eq. (8.6). Thus, for a given value of $R_e$, it may happen that the flow in a large diameter pipe can be laminar since it meets compatibility condition (6.1), while for the same value of $R_e$ a smaller diameter pipe can only convey turbulent flow because condition (6.1) is violated.

## 9. Conclusions

Every fluid that allows sound wave propagation must be capable of storing and releasing elastic energy. Finite speed of propagation requires, moreover, that the fluid should possess a non-vanishing elastic shear modulus. Thus, every fluid that allows wave propagation at a finite speed is capable of opposing shear stress when deformed in shear.

There is a limit, however, to the elastic energy that a material can store per unit volume at a finite temperature. This implies, in particular, a limit to the elastic shear stress that the material can oppose to shearing deformation. This elastic limit is the ultimate shear stress and it is a constitutive property of the material itself. In fluids, the ultimate shear stress is very low, which is why the contribution to motion coming from of the elastic deformation of the fluid can be ignored in most cases. It is the ultimate shear stress, however, which controls the onset of turbulence in a laminar flow.

As occur in brittle solids, a fluid breaks as its stress exceeds the ultimate stress limit. However, at a variance with what happens in solids, the pieces of a broken fluid repair themselves instantly and seamlessly as soon as they come in contact together. Thus, as the shear stress in a laminar fluid flow exceeds the ultimate shear stress, the flow breaks into whirling parts and becomes turbulent, still keeping flowing as a single mass. The broken pieces of fluid cannot be observed, unless the energy liberated in the breaking process is large enough to splash the fluid.

The analysis presented in the paper enables us to predict the critical velocity at which a laminar flow becomes turbulent and to calculate, if necessary, the critical value of the Reynolds number. To do this, the value of the ultimate shear stress of the fluid must be known. The experimental determination of this quantity poses no problems (Sect. 4). Besides giving a rationale to the onset of turbulence, the results of the present analysis can be used to spot the points of the flow where the maximum shearing rate is largest and, thus, evaluate where and when turbulence is bound to appear in the flow. The procedure to do this is quite simple once the velocity field in the non-turbulent regime is known. It should be of help in the design for minimum drag.